\documentclass[twocolumn,epjc3]{svjour3}
\smartqed  % flush right qed marks, e.g. at end of proof
\RequirePackage{graphicx}
\journalname{Eur. Phys. J. C}

\newcommand{\be}{\begin{equation}}
\newcommand{\ee}{\end{equation}}

\newcommand{\no}{\nonumber\\}
\newcommand{\ba}{\begin{eqnarray}}
\newcommand{\ea}{\end{eqnarray}}
\def\bfpi{\mbox{\boldmath $\pi$}}
\usepackage{graphicx}
\usepackage{amsfonts}
\makeatletter
\renewcommand{\@biblabel}[1]{#1.}
\makeatother

\usepackage{geometry}
\begin{document}

\title{Stellar matter with pseudoscalar condensates%\thanksref{t1}
}


\author{A. A. Andrianov\thanksref{e1,addr1,addr2}, V. A. Andrianov\thanksref{e2,addr1}, D. Espriu\thanksref{e3,addr2}\and S. S. Kolevatov\thanksref{e4,addr1}
}

\thankstext{e1}{e-mail: a.andrianov@spbu.ru}
\thankstext{e2}{e-mail: v.andriano@rambler.ru}
\thankstext{e3}{e-mail: espriu@icc.ub.edu}
\thankstext{e4}{e-mail: s.s.kolevatov@gmail.com}


\institute{Saint-Petersburg State University, 7/9 Universitetskaya nab., St.Petersburg, 199034
Russia\label{addr1}
           \and
 Departament d'Estructura i Constituents de la Mat\`eria and
Institut de Ci\`encies del Cosmos (ICCUB),
Universitat de Barcelona,
Mart\'\i \ i Franqu\`es 1, 08028 Barcelona, Catalonia, Spain           \label{addr2}
}

\date{Received: date / Accepted: date}

\maketitle

\begin{abstract}
In this work we consider how the appearance of gradients of pseudoscalar condensates in
dense systems may possibly influence the transport properties of photons
in such a medium as well as other thermodynamic characteristics. We adopt the hypothesis
that in regions where the pseudoscalar density gradient is large the properties of photons
and fermions are governed by the usual lagrangian extended with a Chern-Simons interaction for
photons and a constant axial field for fermions.
We find that these new pieces in the lagrangian produce non-trivial reflection coefficients
both for photons and fermions when entering or leaving a region where the pseudoscalar
has a non-zero gradient.
A varying pseudoscalar density may also lead to instability of some fermion and
boson modes and modify some properties of the Fermi sea. We speculate that
some of these modifications could influence the cooling rate of stellar matter (for instance in compact stars) and have other
observable consequences. While quantitative results may depend on precise astrophysical
details most of the consequences are quite universal and consideration
should be given to this possibility.
\end{abstract}

%\renewcommand{\theenumi}{\arabic{enumi}}
%\renewcommand{\labelenumi}{\arabic{enumi}}
%\renewcommand{\theenumii}{.\arabic{enumii}}
%\renewcommand{\labelenumii}{\arabic{enumi}.\arabic{enumii}.}
%\renewcommand{\theenumiii}{.\arabic{enumiii}}
%\renewcommand{\labelenumiii}{\arabic{enumi}.\arabic{enumii}.\arabic{enumiii}.}

%%%%%%%%%%%%%%%%%%%%%%%%%%%%%%%%%%%%%%%%%%%%%%%
%\vspace{-17cm}
%\begin{flushright} ICCUB-??
%\end{flushright}
%\vspace{15.75cm}
%%%%%%%%%%%%%%%%%%%%%%%%%%%%%%%%%%%%%%%%%%%%%%%

\section{Introduction}
Interest in possible violations of Lorentz and other fundamental symmetries
emerged\cite{LSB} after the paper \cite{Carroll:1989vb}, where electrodynamics with an
additional Chern-Simons (CS) parity-odd term including a constant CS four-vector was considered.
The same authors entertained
the possibility that such an effect could already be
visible in the radiation observed from distant radio galaxies, arriving at a negative
conclusion. Nevertheless, there are some areas where modified Maxwell-Chern\--Simons\- (MCS)
electrodynamics (also known as Carroll-Field\--Jackiw electrodynamics) may be relevant.

It has been suggested that parity breaking phenomena\footnote{
Parity breaking in strong interactions is possible in out-of-equilibrium
processes or with a non-zero chemical potential.}
 may take place in
peripheral heavy ion collisions \cite{cme} manifesting itself in the so-called Chiral Magnetic
Effect (CME)\cite{cme1}. Spontaneous parity violation might also occur  for sufficiently large
values of the baryon density\cite{picon}. Recently several experiments in heavy ion collisions
have also indicated an abnormal yield of lepton pairs\cite{NA60,phenix} and it was conjectured
that the effect may be another manifestation of local parity breaking in colliding
nuclei\cite{aaep,anesp}. This effect would be due to an interplay between
topological fluctuations in QCD (see \cite{moore,Kharzeev:2015kna} and refs. therein) and MCS electrodynamics.

Another source of possible macroscopic manifestations of parity breaking is the presence of an
axion background. Even though there is no direct evidence of the existence of axions yet
it has been speculated that parity odd regions may occur after Bose-Einstein condensation of axion or
axion-like fields \cite{Andrianov:1994qv}--\cite{ArkaniHamed:2004ar}
at several astrophysical scales. In particular this could take place in the interior
of stars\cite{mielke}. Recent investigations speculate that these axion condensates\footnote{
These axion condensates should not be confused with the classical cold axion background --- a viable
candidate for dark matter \cite{raffelt}.} could be
rather compact in size \cite{BEC}.

In dense nuclear matter one should contemplate also phenomena such as neutral pion
conden\-sation\cite{Muto:1988uq} or the occurrence of a disoriented chiral
con\-den\-sate\cite{Mohanty:2005mv}. Both phenomena have received considerable attention in the past.

We will generically refer to these phenomena as pseudoscalar condensation and most
of its consequences will be independent of the precise mechanism triggering the appearance
of local parity breaking.

In fact what matters for our purposes is not the existence of pseudoscalar condensates per se
but rather gradients of the pseudoscalar density. A region with a strictly constant pseudoscalar
density has a vanishing CS four-vector and its electromagnetic properties are described by
usual Maxwell electrodynamics. On the contrary in regions where the pseudoscalar density is
space-dependent a non-zero CS vector appears.  In this region the relevant lagrangian
to describe electromagnetic interactions is the MCS lagrangian.
Similar arguments can be applied to the constant axial-vector field
coupling to fermions\cite{AGS2002}. We shall therefore be interested in this case.

In the next sections we discuss several effects resulting from the appearance of pseudoscalar
gradients in dense systems. We first consider how they may affect the propagating modes
of photons and fermions. For the former, different polarizations are affected differently, while in the
case of fermions the Fermi sea is split in two, with slightly different Fermi levels. We then proceed
to determine transmission and reflection coefficients
for photons crossing a layer where a pseudoscalar condensate has a gradient, concluding that it may
have a relevant influence in process mediated by radiative cooling if such gradients are sufficiently large
compared to the temperature. A similar phenomenon occurs in principle for fermions, but it is much
suppressed due to their larger mass. Fermions can `decay' (i.e.
move from the higher Fermi sea to the lower) in such a medium emitting a photon \cite{AGS2002,Zhuk2006}. Conversely,
photons propagating in such a medium but free otherwise can materialize in a fermion-antifermion pair in
certain circumstances. To conclude we speculate about the possible relevance of the above phenomena
in degenerate fermion systems such as compact stars where, as previously argued pseudoscalars
condensates may be present.

\section{Spectrum of photons in a pseudoscalar condensate gradient}
Let us commence our analysis by assuming that in a dense system  there are domains
with different pseudoscalar densities. These domains may be either of axion or pion-like type. We will
characterize the different domains by approximately constant values of the condensate $a_i$. Obviously
$a=0$, i.e. no parity breaking at all, is also a possibility.
Typically, transition regions between different domains and between a parity breaking
region and the normal vacuum will be present.
The figure shows the generic situation we are considering: region 2 is a transition region with non-zero
pseudoscalar gradient separating regions 1 and 3.
\begin{figure}[ht]
\center{\includegraphics[width=0.5\linewidth]{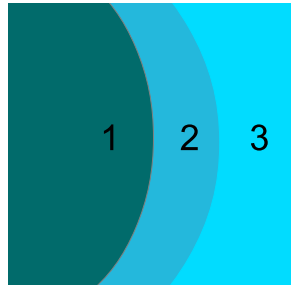}}
\caption{A sketch of possible domains with pseudoscalar condensates. }
\end{figure}
Inside a transition region ---one with a non-zero gradient--- the behavior of the photons can be described
by Carroll-Field-Jackiw model, or
MCS electrodynamics, with the Lagrange  density
\begin{em}
\begin{eqnarray}
{\mathcal L} =  &-&\,{\textstyle\frac14}\,F^{\alpha\beta}(x)F_{\alpha\beta}(x)
-\,{\textstyle\frac14}\,F^{\mu\nu}(x)\widetilde F_{\mu\nu}(x)\,a(x) \nonumber \\
%+ {\textstyle\frac12}\,\zeta_\alpha A_\beta(x)\tilde F^{\,\alpha\beta}(x)
%\nonumber\\
%&+&
&+& {\textstyle\frac12}\,A_\mu(x)m^2_t\,{\cal P}^{\mu\nu} A_\nu(x) \label{lagrangian},
\end{eqnarray}
\end{em}
where $A_\mu(x)$ and $a(x)$ stand for the vector and effective background pseudoscalar fields respectively,
$\widetilde F^{\mu\nu}={\textstyle\frac12}\,
\varepsilon^{\,\mu\nu\rho\sigma}\,F_{\,\rho\sigma}$ is a dual field strength and $m_t$ is an effective transverse photon mass \cite{kapusta,frad}(see Sec.6) which is generated by gauge invariant polarization tensor in plasma, containing the transversal projector  ${\cal P}^{\mu\nu} = \eta^{\mu\nu}- \partial^\mu\partial^\nu / \partial^2 .$  If one uses the low momentum limit  ${\mathbf k} \to 0$, at high temperatures $T$ and chemical potentials $\mu$, this mass happens to be constant, $m_t^2 \simeq \frac{e^2}{3}\left( \frac13 T^2 +\frac{\mu^2}{\pi^2}\right)\not=0.$

The classical background
is different for different regions. In the area outside the boundary layers (regions 1 and 3) $a(x)$ is constant.
In this case the second term in Lagrangian does not give any corrections to field equation and one
recovers standard electrodynamics. The only nontrivial area to describe is the region 2. We assume that in
this layer $a$ changes linearly from $a_-$ to $a_+$ across the gap. Then
inside the region 2 the relevant pseudoscalar background can be locally described by
\begin{eqnarray}
a(x)&=&\,\zeta\cdot x [\theta(\zeta\cdot(x-x_-))-\theta(\zeta\cdot(x-x_+))]  \to \nonumber \\
\zeta\cdot x
&\equiv& \zeta_\mu x^{\mu};  \label{background1}
\end{eqnarray}
in which a fixed constant four vector $\zeta_\mu$ with mass dimension  is used as an argument.
$\zeta_\mu$ is actually proportional to the local gradient of the pseudoscalar
condensate. We assume that the wave function of photons is considerably shorter
that the thickness of the layer and accordingly take the width of the latter to infinity to
simplify our calculations.
In this case Lorentz invariance would be violated in the Minkowski half space.

If we consider a small area that does not feel the curvature of the shell\footnote{Again, this is a good
approximation for photons whose wavelength is much shorter than the
characteristic sizes involved}, we may choose
for simplicity the first coordinate along the local radius of curvature of the
bubble and assume the first boundary $x_-$ to be located at $x_1=0$. In this particular case $(\zeta_\mu) = (0,\zeta,0,0)$,
\begin{equation}
a(x)=\,\zeta x_1\,\theta( x_1),\label{background2}
\end{equation}
where we have now assumed that the thickness of the boundary is much
larger than the characteristic photon wave length and accordingly we have taken $x_+ \to
\infty$. This approximation makes the calculation simpler as it allows to decouple the
effect of the two successive interfaces.

Then the wave equation reads
\begin{eqnarray}
(\Box + m^2_t) A^\nu  + \zeta \varepsilon^{1\nu\sigma\rho}\,\theta( x_1)\partial_\sigma A_\rho = 0
\label{systspat}
\end{eqnarray}
Its solution in the MCS medium can be found by proper projection on the longitudinal and chiral modes
(see \ref{projectors}). The corresponding dispersion relations are
\begin{eqnarray}
\left\lbrace
\begin{array}{l}
k_{1L}^{CS}=k_{1}^0=\sqrt{\omega^2-k_\bot^2};\\
k_{1+}^{CS}=\sqrt{\omega^2-m^2_t-k_\bot^2-\zeta_1 \sqrt{\omega^2-k_\bot^2}};\\
k_{1-}^{CS}=\sqrt{\omega^2-m^2_t-k_\bot^2+\zeta_1 \sqrt{\omega^2-k_\bot^2}} ,
\end{array}\right.\
\label{polarizations}
\end{eqnarray}
where the index `0' labels the medium with the usual  vacuum dispersion relation, $\omega$ is a photon energy and ${\vec k}_\bot = (0,k_2,k_3)$ is a transversal photon wave vector.
This expression describes three different physical polarizations in the MCS medium for $x_1> 0$. Of course
the standard Maxwellian behavior is recovered by setting $\zeta=0$ in all expressions, which
corresponds to the region $x_1<0$, where there is no Lorentz symmetry breaking.

\section{Spectrum of fermions in a pseudoscalar condensate gradient}
In this section we outline the key points of the properties
in 3+1 Minkowski space-time of
free spinor fields in the presence of a Lorentz covariance
breaking kinetic term associated with a constant axial-vector $b_\mu$.
This vector is supposed to be generated in the region  between two bubbles with different pseudoscalar
condensate and its presence will change drastically some fermion properties
in the vicinity of Fermi surface.

The free fermion spectrum can be obtained from a modified
Dirac equation in the momentum representation
\begin{equation}
\left(\gamma^\mu p_\mu  -m -\gamma^\mu b_\mu\gamma_5\right)\psi = 0\,.
\label{3.1}
\end{equation}
The solution of this equation is discussed \\in \ref{appDiracEq}. The fermion
spectrum is determined by the on-shell condition
\begin{equation}
\left(p^2 + b^2 - m^2\right)^2 + 4 b^2 m^2 - 4 (b \cdot p)^2 = 0\,.
\label{3.2}
\end{equation}
This equation has real solutions for any
value of $b_\mu$.  However, a consistent quantization of the
spinor field can be
performed if there are two pairs of opposite-sign roots of eq.~(\ref{3.2})
and a mass gap between them.  This condition holds true~\cite{6} for
sufficiently small $b_\mu$ and a
mass gap between positive and negative frequencies exists provided
that no solutions with $p_0 = 0$ exist. Such solutions never arise for space-like $b_\mu$ which we employ in this paper.
Moreover the choice ${\vec b}^2 < m^2$ is quite plausible as the magnitude of the constant axial vector associated
to a gradient of pseudoscalar condensate is expected to be of the order of a few keV at most (see
discussion in next section).

For the purely space-like case one can fix $b_\mu =
(0,\vec b)= (0,b,0,0);\, b>0 $ by the  proper choice of coordinate system\footnote{In concordant frames the results must be modified but the physical consequences remain the same \cite{shabad}}. Then the dispersion law is given by
\begin{eqnarray}
\omega^2 &=& \vec p^2_\perp +p_1^2+ b^2 + m^2
\pm 2 b \sqrt{p^2_1 + m^2}\no &=&  \vec p^2_\perp +\left(b \pm \sqrt{p^2_1 + m^2}\right)^2;\no
\vec p &=& (p_1, p_2,p_3);\quad \vec p_\perp = (0, p_2,p_3) .
\label{3.6}
\end{eqnarray}
These solutions are separated by the stability cone. The stability border $p_\mu^2 = 0$ is described by
\begin{equation}
|p_1| = \frac{(m^2 -b^2)}{2 b}\,.
\label{3.7}
\end{equation}

\section{Fermi sea in a pseudoscalar condensate gradient}
In this section we shall assume that the momenta involved fulfill the condition
$p \ll m$ and therefore a non-relativistic approximation is valid.

Taking this into account the dispersion relations for fermions in
a medium with a non-vanishing pseudoscalar gradient would approximately read
\begin{equation}
\omega= m \pm b + \frac{p_1^2}{2m} +\frac{p_\perp^2}{2m}.
\end{equation}
 Let us assume that the medium is in thermal equilibrium with a temperature $T$. The two different mass-shell conditions established in the
previous section will give rise to two different Fermi seas. These
in the non-relativistic limit correspond to the $\pm$ branches in the above
equation. Assuming that initially N+=N-=N the corresponding chemical potentials $\mu_\pm$ are obtained by solving eqs. ,
\begin{equation}
N = V \int d\omega \, \frac{\sqrt{(2m)^3(\omega\pm b)}}{4\pi^2}\, \frac{1}{\exp{\frac{\omega-\mu_\pm}{T}}+1}
\end{equation}
 in respect to $\mu$. Clearly this gives $\mu_+ = \mu_- + 2b$ and therefore $\epsilon_F^+ > \epsilon_F^-$ (we assume
that $b$ is positive; otherwise the situation is reversed).
The existence of pseudoscalar regions with a non-zero gradient thus naturally lead to a splitting of the
Fermi sea in two, with a difference in levels directly proportional to the gradient itself. For simplicity
we are implicitly assuming that the gradient is constant (i.e. the pseudoscalar density varies
linearly along the `1' direction).

At least theoretically one could contemplate a single pseudoscalar condensate occupying a large part of
a star primarily made of degenerate fermions (such a neutron star or a white dwarf) in its
interior.
Let us also assume for the sake of the argument that the condensate varies linearly with the radius.
From the previous considerations this would
mean that the degenerate electrons or neutrons that sustain the star from collapsing, in fight against
the gravitational potential, are split. Half of the fermions (the `+' ones) have to deal with more than
their fair share of the weight of the star. The natural tendency under the action of gravity would then be to
equilibrate both Fermi seas. In order to do so, some of the `+' fermions have to turn into `-' ones. Is this possible?
The answer is yes and it will be discussed in the coming sections. The conversion would
necessarily imply a
large emission of photons with energies up to $\omega \simeq 2b$.

The same phenomenon can present itself in smaller regions of the star leading to local imbalances in the Fermi sea
that could trigger small starquakes with their corresponding photon emissions.

\section{Propagation of photons across pseudoscalar gradients}
Now we describe the propagation of photons inside every of three regions indicated in Fig. 1.
Regions 1 and 3 correspond to taking $\zeta=0$ while $\zeta\neq 0$ in the
intermediate transition region 2 where the pseudoscalar condensate has a gradient. In this section we neglect an transverse photon mass $m_t$.
As it is manifest in the previous equations, different regions
lead to different dispersion relations. As a consequence non-trivial reflection and
transmission coefficients between the different regions appear. This issue was first discussed
in \cite{AK}.

\subsection{Entering the boundary layer}
We assume that the thickness of the intermediate shell
(region 2 in Fig. 1) is much larger than a typical wavelength and a mean free path.
Under these assumptions (\ref{systspat}) works.
Photons fall on the boundary from a region where ordinary electrodynamics holds and attempt to penetrate
in one governed by MCS electrodynamics. Matching conditions for this problem were discussed in
detail in \cite{Andrianov:2011wj}. To understand which photons penetrates into the shell it is worth
to recall the discussion. Solution of (\ref{systspat}) can be found by using Fourier transformation over all
components but $x_1$ and may be written in the form

\ba
&&\tilde A_\nu = \\&& \left\lbrace
\begin{array}{l}
\tilde u_{\nu \rightarrow}e^{ik_{1}^0 x_1}+\tilde u_{\nu \leftarrow}e^{-ik_{1}^0 x_1},\!\ x_1<0; \\ \\
\sum\limits_A \left[\tilde v_{\nu A \rightarrow} e^{ik_{1A}^{CS} x_1}+
\tilde v_{\nu A \leftarrow} e^{-ik_{1A}^{CS} x_1}\right],
 \!\  x_1>0, \nonumber
\end{array} \right.
\label{anutilde}
\end{eqnarray}
where $\tilde v$ and $\tilde u$ depend on $(\omega, k_2,k_3)$. The first index of $\tilde v$ denotes  the corresponding component of $A_\nu$, $\nu=0,2,3$,
the second index $A$ stands for the polarizations $L, +, -$ and the arrows $\rightarrow$, $\leftarrow$
point out the direction of particle propagation. It is necessary to distinguish among the various polarizations
because they obey different dispersion relations. Note that we contemplate the possibility of photons
developing a transverse mass $m_t$ which can
be easily reinstated in the formulae.

Then we take the initial amplitude, and using eqs. (\ref{u0}-\ref{tr}) in \ref{transmission}
we find the amplitudes of
the transmitted  waves with polarizations $L, +, -$
\ba
{\mathfrak T}^{\pm}=\frac{1}{1+\sqrt{1\mp\frac{\zeta}{k_1^0}}}. \label{coeff}
\ea
Longitudinally polarized photons are not
affected by the change in the medium.
\begin{figure}[ht]
\center{\includegraphics[width=0.8\linewidth]{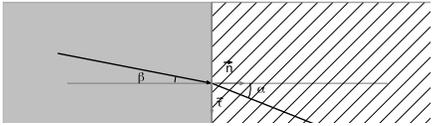}}
\caption{The geometry of photon propagation. $\vec n$ is a normal vector. In the left region (region 1) the
pseudoscalar condensate takes a constant value. Region 2 (right) is assumed to be governed by
MCS electrodynamics describing a varying pseudoscalar condensate as befitting a transition region} %\\{\mbox{\rm-LаШб. 1}}}-A
\end{figure}

The direction of outgoing photons corresponds to the angle $\beta$ (Fig. 2). After the propagation
through the boundary, their direction will be\\ changed (angle $\alpha$) accordingly to their polarization.
We decompose $\vec k = k_n \vec n + k_\bot \vec \tau$ (Fig. 2).
From \cite{Andrianov:2011wj} we know that $k_\bot$ remains the same after crossing the boundary.
This means that the trajectory of the particle and normal vector lie in one plane. However, $k_n$ changes.
We introduce the new variable $k_n^{CS}$ to describe the normal component of $\vec k$
after crossing the boundary. For $x_1<0$ one has
\ba
\cos \beta = \frac{k_1^0}{\omega}
\ea
We are interested in finding the overall flux of outgoing photons from region 1 to region 2 in Fig. 2.
To do this we first consider a small volume near the boundary. We assume that it radiates uniformly
in all directions. For us is important the flux of energy which propagates outwards so we are interested
only in upper half-sphere in Fig. 3.
\begin{figure}[ht]
\center{\includegraphics[width=0.8\linewidth]{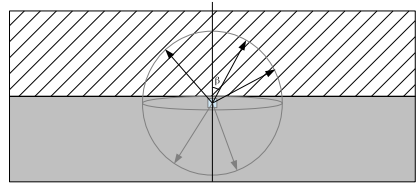}}
\caption{Radiation of a layer}
\end{figure}
We assume that this small volume radiates with a certain frequency $N^\omega_\Omega$ per solid angle. At this moment we neglect the photon mass to get some quantitative results.
In order to find the total luminosity we should integrate over the solid angle, frequency and surface of the layer.
The last integration would be the same for the case with or without pseudoscalar condensation
\ba
L &\propto& \int^\infty_{0} d \omega N^\omega_\Omega  \int_{\min (\frac{\pi}{2},\arccos (\frac{\zeta}{\omega}))}^{\frac{\pi}{2}} d
\beta {\mathfrak T}^{+}(\beta, \omega, \zeta) \nonumber\\
&+&  \int_0^{\infty} d \omega N^\omega_\Omega
\int_0^{\frac{\pi}{2}} d \beta {\mathfrak T}^{-} (\beta, \omega, \zeta)
\ea
Here one can see that integration over angles in the first term begins from the value
$\cos \beta = \frac{\zeta}{\omega}$. This value comes from kinematical condition of the positive polarization.
It is easy to see from (\ref{coeff}), that for ${\mathfrak T}^{+}$ in denominator we have a negative
value under the root if $\zeta > k_1$. Physically it means, that for falling photons with $k_1 < \zeta$ it
is kinematically forbidden to convert into positive polarization photons in the medium with a linearly
varying pseudoscalar field. For negative polarization there is no restriction, in
 (\ref{polarizations}) $k_{1-}$ is positive for any values of $\zeta$ and as a result, we do not
see any special limits of integral for negative polarization.

Next we assume that the medium is in thermal equilibrium with a temperature
$T$ and accordingly $N^\omega_\Omega \propto \omega^3 / (e^{\frac{\omega}{T}}-1)$.
We will compare this value with the luminosity of the same volume without any parity breaking
boundary effect. Let us call this last value $L_{0}$
\ba
L_{0} \propto \int_0^{\infty} d \omega N^\omega_\Omega \int_0^{\frac{\pi}{2}} d \beta.
\ea
Finally, we can plot a graph which demonstrate the effect of the intermediate shell on the luminosity.
\begin{figure}[ht]
\center{\includegraphics[width=0.9\linewidth]{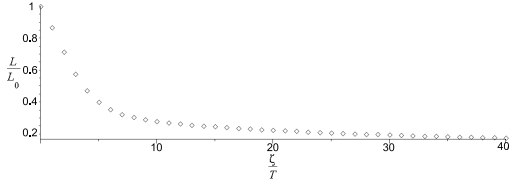}}
\caption{The outgoing energy flow ratio for the two cases (with and without
 pseudoscalar condensate)}
\end{figure}
From Fig. 4 one can see that the effect of changing the dispersion relation
across the boundary between region 1 with $a=$ constant and region 2 where $a$ depends
linearly in $x_1$ is very noticeable. At large values of $\zeta$ (compared with temperature)
most photons are reflected from the boundary and accordingly the energy radiated across the boundary
decreases. Note that as a consequence the radiation emitted does not correspond exactly to that
of a black body with temperature $T$.

\subsection{Escaping from the boundary layer}
After escaping the first region photons appear, if they are not reflected, in the
intermediate shell where MCS electrodynamics is at work. To leave this medium and gain access to
another domain where $a=$ constant (possibly zero)
photons have to pass through one more boundary. This corresponds to
the boundary between regions 2 and 3 in Fig. 1. We use the same technique as in the previous section.
Figs. 2 and 3 still apply but reversing the areas where ordinary and MCS electrodynamics apply.
In this case
\ba
\frac{k_n^{CS}}{k_\bot}=\cot(\alpha); \qquad \qquad \frac{k_n}{k_\bot}=\cot(\beta). \label{angles}
\ea
Furthermore we know that for spatial CS vector there are two transversal polarizations in the
MCS  medium with the dispersion relations
\ba
k_{n\pm}^{CS}=\sqrt{\omega^2-k_\bot^2\mp\zeta \sqrt{\omega^2-k_\bot^2}}
\ea
or, since we consider photons and $k_{n}=\sqrt{\omega^2-k_\bot^2}$,
\ba
k_{n\pm}^{CS}=\sqrt{k_n^2\mp\zeta k_n};
\ea
and
\ba
\frac{k_n}{k_\bot}=\cot(\alpha); \qquad \frac{\sqrt{k_n^2\mp\zeta k_n}}{k_\bot}=\cot(\beta).
\ea
Using the results obtained in \cite{Andrianov:2011wj}, one may find the transmission coefficient
of outgoing particles for every polarization
\ba
{\mathfrak T}^{A}=\frac{2 k_{1A}^{CS}}{k_{1A}^{CS}+k_{1}^0}.
\ea
We are interested only in transversal polarizations (we deal with photons, however, the longitudinal
one does not feel the boundary anyway), so in our terms we write,
\ba
{\mathfrak T}^{\pm}=\frac{2 k_{n\pm}^{CS}}{k_{n\pm}^{CS} + k_n}=\frac{2 \cot(\alpha)}{\cot{\alpha}+\cot(\beta)}.
\ea
For our purposes it is necessary to express  ${\mathfrak T}^{\pm}$ as a function of $\beta$ that means
expressing $\alpha$ in terms of $\beta$. For constant $a(x)$,  $\omega=|\vec k|$ and one can
use $k_n=\omega \cos \alpha$. So, (\ref{angles}) gives
\ba
\cot \alpha = \cot \beta \frac{\omega \cos \alpha}{\sqrt{\omega^2 \cos^2 \alpha \mp \zeta \omega \cos \alpha}}.
\ea
Where as usual the $\mp$ stands for different polarizations.

Solving this equation one can find the expression for $\cot \beta$ for different polarization
and the value of  ${\mathfrak T}^{\pm}$,
\ba
{\mathfrak T}^{\pm} (\beta, \zeta, \omega)=
\ea
{\small
\ba
\frac{2 \cot \beta}{\cot \beta + \frac{\pm \zeta
+ \sqrt{\zeta^2+4\omega^2 \cot^2\beta (1+\cot^2 \beta)}}{\sqrt{4\omega^2(1+\cot^2 \beta)-2\zeta^2\mp2\zeta\sqrt{\zeta^2
+4\omega^2 \cot^2 \beta (1+ \cot^2 \beta)}}}} \nonumber
\ea}
Using this formula we may find for any angle $\beta$ the fraction of incoming photons succeeding in
crossing the boundary at $x_+$, i.e. the boundary between the regions 2 and 3 in Fig. 1.
Like in the previous section we consider a small volume which radiates at certain frequency $N^\omega_\Omega$ in a
unit of solid angle and write total luminosity
\ba
L &\propto& \int^\infty_\zeta d \omega N^\omega_\Omega  \int_0^{\frac{\pi}{2}} d \beta {\mathfrak T}^{+}(\beta, \omega, \zeta) \nonumber \\
&+&  \int_0^{\infty} d \omega N^\omega_\Omega \int_0^{\frac{\pi}{2}} d \beta {\mathfrak T}^{-}(\beta, \omega, \zeta)
\ea
It is worth commenting on the integration region of both terms. Looking at the expression (\ref{angles})
one may see that for negative polarization there are no any restriction on the kinematics of photon.
This fact means we should integrate over all energies of outgoing photons.
In case of positive polarization $\omega$ cannot be less than $\zeta$.

Like in previous section, to show the qualitative effect of pseudoscalar condensate we assume that
the medium has a temperature $T$.
\begin{figure}[ht]
\center{\includegraphics[width=0.9\linewidth]{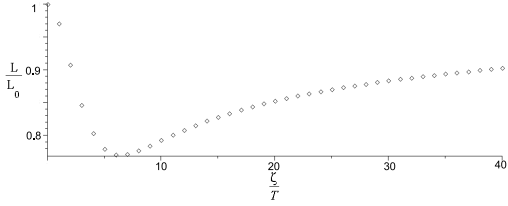}}
\caption{The outgoing energy flow ratio for the two cases
(with and without pseudoscalar condensate)}
\end{figure}
From Fig. 5 one can see that for $\zeta>T$ there is a clear effect on the flux of outgoing energy. However, the boundary between regions 2 and 3 (Fig. 1) does not play a crucial role in the decreasing of energy flux effect (compare Fig. 4 and Fig. 5). Nevertheless, to get the total effect, one should take into account the impact of both boundaries.
\begin{figure}[ht]
\center{\includegraphics[width=0.9\linewidth]{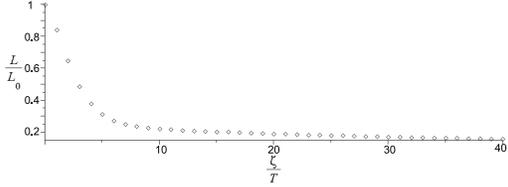}}
\caption{The outgoing energy flow ratio calculated for both boundaries}
\label{coolingTohether}
\end{figure}
If we neglect the thermal capacity of the shell \footnote{This implies assuming that the thickness
of the shell is much smaller
linear sizes of a typical domain where $a=$ constant.} then we may plot a graph
(Fig. \ref{coolingTohether}) showing the combined effect of two boundaries of the domain.
One sees that the effect may be substantial and in this case the thermal
evolution of the bulk would be affected: radiative cooling (assuming that region 1 is hotter than region 3)
would proceed more slowly. In systems like neutron stars or white dwarfs radiative cooling is not
the dominant cooling mechanism and therefore the impact of the previous effect is limited.

\section{Fermion decay in a pseudoscalar gradient}
Fermions crossing domains with a pseudoscalar gradient will reveal the same effects of
partial reflection as the photons show for exactly the same reasons.
The discussion of these effects is given in \ref{boundaryElectrons}.
However if $b\ll m$ then the effects are clearly  subleading.

A more important phenomenon is the instability of fermions in a parity breaking environment.
Namely, a fermion (be it an electron or a neutron) with a larger effective transversal
mass $m_+ = \sqrt{p_1^2 +m^2} +b$ may well decay into a fermion with a smaller effective
transversal mass\\ $m_- = \sqrt{q_1^2 +m^2} -b$ by emitting
a photon (in the case of electrons) or a neutral virtual pion (in the case of neutrons) that
would immediately decay into two photons. Since the formalism is identical
in both cases let us restrict ourselves to the case of photon emission via this mechanism for
distorted electrons. In this case the emergent photon will have a transversal
mass $m_\perp = \sqrt{k_1^2 + m^2_t +\zeta^2/4} \pm \zeta/2$
where $m_t$ is the effective transversal photon mass in the plasma (if any).

Let us restrict ourselves to the purely
space-like $b_\mu= (0,b,0,0)$ and take $b > 0$. Besides one can use the remaining small Lorentz
invariance $O(1,2)$ in the hyperplane $(0,2,3)$ and select out the transversal rest frame for
the decaying electron.
Then the decay is possible for an electron with transversal mass $m_+$ and momentum $p_\mu$
into an electron with transversal mass $m_-$ and momentum $q_\mu$  and a photon with
momentum $k_\mu$ and transversal mass $m_\perp$
.
Ener\-gy\--mo\-mentum balance for the decay, $p_\mu = q_\mu + k_\mu$, consists of the following relations
\ba
b&+& \sqrt{p_1^2 +m^2} = \sqrt{\vec q^2_\perp + (\sqrt{q_1^2 +m^2} - b)^2}+ \no
 &+& \sqrt{\vec k^2_\perp+
(\sqrt{k_1^2 + m^2_t +\varepsilon^2\zeta^2/4} +\varepsilon \zeta/2)^2};\no
p_1 &=& q_1 + k_1;\quad \vec q_\perp = - \vec k_\perp ,
\ea
where $\varepsilon = 0,\pm 1$ for photon longitudinal and two $\zeta$ polarizations.

In order to estimate the threshold of the reaction that allows `+' electrons to decay
into `-' electrons, we neglect the positive contributions of the transverse momenta in the right-hand side,
\ba
b&+& \sqrt{p_1^2 +m^2} > \sqrt{q_1^2 +m^2} - b+ \no
 &+& \sqrt{k_1^2 + m^2_t +\varepsilon^2\zeta^2/4} +\varepsilon \zeta/2,
\ea
it is assumed that $m > b$. Thus,
{\small
\ba
2 b + \sqrt{p_1^2 +m^2}- \varepsilon \zeta/2
> m
+\sqrt{m^2_t +\varepsilon^2\zeta^2/4},
\ea}
wherefrom one derives the necessary inequalities to allow the decay to a particularly polarized photon
\be
\begin{array}{|c|c|l|} L& \varepsilon=0& 2b+\sqrt{p_1^2 +m^2} - m > m_t;\\
+ &  \varepsilon = +1& (2b+\sqrt{p_1^2 +m^2} - m)\times\\&& \times\sqrt{1 - \frac{\zeta}{2b+\sqrt{p_1^2 +m^2} - m}} > m_t ;\\
- &  \varepsilon = -1& (2b+\sqrt{p_1^2 +m^2} - m)\times\\&& \times \sqrt{1 + \frac{\zeta}{2b+\sqrt{p_1^2 +m^2} - m}} > m_t;
\end{array}
\ee
Evidently the absolute bound for the decay threshold is reached in the rest frame of a fermion $p_1 = 0$.
At very high temperatures $T$ and densities (chemical potentials $\mu$) and moderate photon momenta the effective photon
mass\cite{kapusta} $m_t^2 \simeq \frac{e^2}{3}\left( \frac13 T^2 +\frac{\mu^2}{\pi^2}\right)$ suppresses
the transformation of `+' electrons into `-' electrons in the parity breaking layer.

The transformation from one type of fermions to the other via photon emission is the process that
may allow chemical equilibrium between the respective Fermi levels, eventually leading to a small excess
of `-' fermions in the system\footnote{
The possible backreaction of this effect on the value of the pseudoscalar condensate
itself has not been considered.}. As we have seen the viability of the process and the characteristics
of the photon emitted depends crucially on the relation between the three parameters
$b,\zeta$ and $m_t$. If the transition turns out to be energetically impossible because
$m_t$ is too large, in the presence of this phenomenon a star
has to be supported by a combination of the two Fermi seas. Notice that while in vacuum all
`+' electrons would like to decay to `-' electrons, this is only possible for a limited number
of electrons in the Fermi sea, as the new `-' would have nowhere to go all levels being occupied. Therefore
the contribution of this anomalous process to star cooling is very small in absolute terms although
the sudden bursts of photons that may produce could represent a clear observational signal. Whether
this is the case or not depends on the transparency of the crust to these photons. If transmission is
negligible they would only contribute to a slight increase of the star temperature. It is to be expected
though that a sudden reordering of the Fermi levels could produce some sort of glitch if the
neutron star turns out to be a pulsar.

\section{Photon decay}
Finally let us discuss another phenomena, which may give contribution to the flux of outgoing particles, namely,
the possibility of high-energy photon decay in a volume where $a\neq$ constant.
For a star thermodynamic balance the importance
of this phenomenon is probably irrelevant but it is interesting on its own nevertheless.

It was shown in \cite{Andrianov:2009tp} that a photon of positive polarization may decay in presence
of a gradient of the pseudoscalar field into an  $e+e-$ pair. This process will suppress the
number of outgoing photons with positive chirality and possibly increase the number of outgoing electrons,
positrons and antineutrino due to the process
\ba
e^+ + n \rightarrow p^+ + \bar \nu_e.
\ea
We will present here some calculations to understand what may be the quantitative effect of the photon decay.
The total decay width for high-energy photons with positive polarization in
a linearly varying pseudoscalar background is \cite{Andrianov:2009tp}
\ba
\Gamma_+  \simeq \frac{\alpha \zeta}{3}
\ea
In order to evaluate an effect of such decays we use the same model, as in previous section. We assume a
layer where the CS vector is pointed along the radius. We also assume
that the total flux of positive polarized photons outgoing from the layer is $N_0(\omega)$. After photons
with positive polarization propagate inside the region governed by MCS electrodynamics
 their number should decrease as
\ba
N(t,w)=N_0 e^{-\frac{\Gamma_+ \cdot t}{\gamma}},
\ea
where $\gamma=\frac{1}{\sqrt{1-v^2}}$ stays for the Lorenz factor of the particle.
In this section we are going to find an order of magnitude of the decaying process. For the simplicity let us consider photons moving along the pseudoscalar gradient. Dispersion law for these particles is
\ba
\omega
=\sqrt{{\bf k_+}^2 + \frac14 \zeta} + \frac12 \zeta, \nonumber
\ea
from eqs.(\ref{polarizations}). We have neglected here a plasma effective photon mass $m_t$.

Now we will find the $\gamma$ factor. The group velocity reads,
\ba
v = \frac{d \omega}{d k}=\frac{\sqrt{\omega^2-\zeta \omega}} {\omega-\frac12 \zeta} < 1;\\
\gamma = \frac{1} {\sqrt{1-v^2}}=\frac{2\omega-\zeta}{\zeta}.
\ea
And we get
\ba
N(t,\omega)=N_0 e^{-\frac{t\cdot \alpha \zeta^2}{3(2\omega-\zeta)}}
\ea
It is important to remind that there is a threshold for the described decay
\ba
\omega \simeq \frac{m_e^2}{\zeta}.
\ea
This may be a large suppression factor for photons of positive chirality. Let us assume for the
sake of discussion that photon energies are of the order $\omega \sim$ GeV and $\zeta$ is
$\geq$ keV. Then one
can easily see that for $t > 10^{-9} s$ most of photons with positive chirality have decayed.
This time ($10^{-9} s$) corresponds to a distance scale $\sim 10  cm$, which is a quite small number for astrophysical object.
So we have to conclude that if inside the medium there is a strong enough gradient of pseudoscalar background,
photons with positive polarization will decay in $e^+e^-$ giving increasing the lepton pairs going and
suppressing almost completely the number of photons with positive chirality (of course reversing the sign
of the gradient the same argument applies to the opposite chirality).

For the process to be possible in a dense fermion medium the resulting fermions must have energies
above the respective Fermi sea levels (recall that they are slightly different for the two fermion
`species').

\section{Possible implications}
From the previous discussion we have seen that there are new potentially interesting effects taking place
in a medium where a pseudoscalar condensate with a non-zero gradient is present. Whether they are
relevant for real physical systems, possibly in compact stars, depends on different circumstances. Some of them
can be assessed on general grounds but others depend on precise astrophysical details.

Before entering into the discussion it may be interesting to understand the magnitude of the relevant
parameters $b$ and $\zeta$.

A hypothetic axion condensate related to
an axion of the Peccei-Quinn type may well couple to electrons although its value is
largely arbitrary but expected to be of order $y_e\times \frac{\langle \partial_xa(x)\rangle}{f_a}$ where
$y_e$ is the corresponding Yukawa coupling and $f_a$ the axion decay constant. $a(x)$ is the varying
axion field. For neutrons the corresponding
coupling would be enhanced by a factor $\simeq 20\div 30$.  The main uncertainty here is
the value to assume for $\langle \partial_xa(x)\rangle$. This quantity also determines $\zeta$.

A pion condensate could make influence on electrons with a coupling of $10^{-3}$ less than for photons
which is related to the $\pi \to e^+e^-$ decay\cite{pdg}. We follow the paper \cite{AGS2002} and refer
the reader to it for more details. As for neutrons, the influence of a varying pion condensate
would be substantially larger and it can be estimated to be of order
\begin{equation}
b\simeq g_{\pi NN} \frac{\langle \partial_x\pi(x)\rangle}{f_\pi}.
\end{equation}
Thus $b$ is potentially large in such a situation.

Well known estimates indicate that the Fermi energy is of the order of $0.10$ MeV
for the degenerate electrons inside a white dwarf and of the order of $50$ MeV for the degenerate neutrons in a typical neutron
star. Clearly in both cases $\epsilon_F << m$ and therefore a non-relativistic approximation
such as the one used to establish the energy levels is reasonable.

Possibly the clearest effect predicted is the splitting of the Fermi sea in two separate sets
corresponding to  different fermion `species' that obey different dispersion relations. In the
non-relativistic approximation the two Fermi levels are separated by the quantity $2b$. Under
sufficient gravitational pressure the two levels will tend to equilibrate with a subsequent
copious emission of photons with characteristic frequency $\sim b$. Depending on specific details
these sudden bursts could correspond to low frequency radio waves. The qualitative estimations of the rate of this interesting process will be done elsewhere.

The possible consequences due to the change in the transmission and reflection properties of photons
are harder to estimate. In stars it takes a long time for radiation to reach surface.
For instance, in the Sun the photon diffusion time scale is $\sim10^5$ years \cite{Sun}.
Neutron star are much denser ($\rho \sim \rho_0$),
so the mean free path of photons is very small (a full picture about the structure of neutron stars can be
found in \cite{NeutronStar}) and it is accepted that, after an initial stage dominated by
neutrino emission, during most of the life of the star heat transfer proceeds through thermal diffusion
in the highly degenerated neutron gas and radiative cooling is only effective
in the outer layers of the star (crust and atmosphere). The internal temperature is nearly constant
thanks to the high thermal conductivity of the neutrons, a consequence of their extremely large mean
free paths as befits a nearly degenerate Fermi gas. The
actual rate of cooling of a neutron star depends very substantially
on the opacity of the outer layers (`heat blanketing envelope' \cite{NeutronStar}).
Thus it seems that the mechanisms discussed in this work could be relevant for cooling if large
pseudoscalar gradients are present near the star's crust, assuming of course that the commonly
accepted cooling mechanisms are correct.

White dwarfs are constituted by a highly degenerated relativistic electron gas and the mechanism of cooling
in their quasi-steady state proceeds very similarly \cite{WD}.
Electron-electron interactions are clearly subdominant
for same reason as neutrons in a neutron star, and transport is dominated by electron-ion
interactions (ions can be treated classically as the temperature is comparable to their chemical
potential, while $T\ll \mu_e$). Not only that, actually most of the white dwarf heat is stored
in the positive ions (recall that for the ions classically $C_V^i \sim \frac32 N_i$, whereas for
the electron degenerate Fermi gas\\ $C_V^e \sim N_ e T/\epsilon_F \ll C_V^i$). Photons in the few keV range
do not interact easily with electrons due to the quantum degeneracy of the latter
having a Fermi energy of the order of $0.10-0.20$ MeV but they do
interact easily with ions and radiative cooling is mostly relevant in the external layers of the star.

In any case the mechanisms discussed here work in the direction of retarding the process of radiative
cooling inasmuch as the latter is relevant. Not only a large fraction of the photons may be reflected;
half of them may actually be unstable if the right conditions are given and may rapidly decay to
fermion-antifermion pairs helping populate the electron sea.

We have considered other possible influences of the distorted dispersion relations, such as changes in
the fermion conductivity, but they are too small to be taken into account.

\section{Conclusions}
In this work we have investigated how the appearance of a pseudoscalar condensate may change the
properties of photons and fermions inside a stellar matter.

While the formal aspects of the work presented here are well founded,
the practical relevance of the present study hinges on a number of hypothesis. In particular, we have
to assume that parity may be spontaneously broken due to the high density, to the presence of some axion
condensate, or both.
We have to accept that several domains of these
characteristics form in the central part of the star, if our model pretend to make a contribution to the compact star inner processes. At the very least there should be one domain
surrounded by an external crust where parity is not broken.

We also have to assume that the characteristic scale
of these domains and also the intermediate transition regions are much larger than typical wave length
of photons (and electrons) present in the star. This last assumption seems guaranteed.The paper considers the situation when the pseudoscalar profile is given by two constant values at  $x<x_-$  and  $x>x_+$  with a  linear interpolation in the intermediate region. Certainly it would be more realistic to consider a setup where the gradient varies smoothly at length scales much longer than the photon / fermion wavelength. In this case it can be interpolated in piecewise line approximation, i.e. the result could be found by combining several domains with constant CS vector in the spirit of Fig. 1 and corresponding convolution of photon reflection/transmission coefficients. We postpone this interesting case  to a forthcoming paper.

The appearance of a pseudoscalar condensate in nuclear matter at high baryon densities has not yet been observed
but it is predicted theoretically in a solid way \cite{anesp}. On the other hand different authors have speculated
with the existence of axion condensates (of a Bose-Einstein type) with radii ranging from 10 km to $10^4$ or
$10^5$ km. These are very relevant sizes for stellar physics.

Then if these hypothesis hold the mechanism of stabilization of the star and even the very
process of cooling can be affected by the presence of a
CS vector induced by a varying pseudoscalar condensate. There is little or none model dependence in the
predictions of this phenomenon. The
transition regions are described by Maxwell-Chern-Simons electrodynamics and their consequences can be worked
out independently of the microscopic details of pion/axion condensation. However in order
to get a numerical estimate of the modified cooling rate the present mechanism would require
a knowledge of the distribution of the different domains, at least in average, and more importantly
an estimate of the value of the pseudoscalar gradients relevant in the present context
of stellar plasmas.

Among the effects that we have predicted we list the following: (a) Photons and fermions get a distorted spectrum. The
former have non-trivial reflection coefficients by layers where the pseudoscalar density, whatever its origin may be,
varies. Inasmuch as radiative processes are relevant, i.e. particularly in regions close to the surface
of the star if a substantial pseudoscalar gradient is present, they can be severely affected retarding the cooling
of the star. (b) A similar effect is present for fermions helping them to build up larger densities in the star's inner
regions but its relevance is much smaller than for photons. However, one half of the fermions are found to be unstable
in the presence of the axial background and can potentially decay with the emission of a photon. Only a small fraction
of fermions can actually decay due to Pauli's exclusion principle. This would have the opposite effect of helping
to cool the star to some extent (radiative cooling is inefficient in the inner parts of a compact star). (c) A
mismatch between the Fermi seas corresponding to the two type of fermions may be suddenly rearranged due to the
star gravitational pressure causing photon emission with a well determined spectrum (Fast Radio Bursts? \cite{FRB}). Needless to say the practical
relevance of these effects depends mostly on the magnitude of the parity breaking parameters $\zeta$ and $b$.

Of course if parity is broken inside a star other consequences should follow since this
would undoubtedly modify the equation of state. There could also be photon birefringence \cite{Andrianov:2011wj}
at the boun\-dary layer and photon instability \cite{Andrianov:2009tp}. All taken together
could help to detect parity breaking in dense (or not so dense) stars.

If focusing particularly on neutron stars the
models should describe the cooling rate of
these objects. In \cite{Alford:1999pb} a predicted decay time was
obtained from a model of color superconductivity and in
a recent paper \cite{Horowitz:2014xca} it was shown how the crust cooling may depend on the presence
and properties of nuclear matter. In \cite{Klochkov:2014ola}, \cite{Homan:2014tba} one may find the review of
differences between the theoretical predictions and experimental data. All in all, it seems fair to conclude that
we still do not have a good understanding of the cooling of neutron stars.

In the near future a number of new astrophysical instruments will provide various information on
neutron stars (see the recent  review \cite{nsnews}). In particular, the Neutron Star
Interior Composition Explorer (NICER) \cite{nicer} to be launched in 2016 is expected to discover
tens of thousands of neutron stars and help understanding their core phenomena and to unravel
footprints of parity breaking due
to pseudoscalar condensation.

\section*{Acknowledgements}
D.E. would like to thank the warm hospitality extended to him at the State University of Saint
Petersburg and at the Perimeter Institute where this work was completed.
We acknowledge the financial support from research projects FPA2013-46570, 2014-SGR-104
and Consolider CPAN. Funding was also partially provided by the Spanish MINECO under project
MDM-2014-0369 of ICCUB (Unidad de Excelencia `Maria de Maeztu')
A.A.,V.A. and S.K. have been also supported by Grant RFBR project 16-02-00348. As
well  A.A., V.A. and S.K. were supported by the Saint Petersburg State University
Grants 11.42.1364.2015 and 11.42.1366.2015 as well as by the Grants\\ 11.41.777.2015 (V.A.) and 11.41.821.2015 (S.K.). S.K. also acknowledges the financial support
from the non-profit Dynasty Foundation. The authors would like to thank F.J. Llanes-Estrada for a critical reading of the manuscript and A. Reisenegger for useful remarks.

\appendix
\section{Solutions of MCS wave equation} \label{projectors}
We build polarization vectors using the projector on the plane
transverse vectors $k_\mu, \zeta_\nu$ \cite{AACGS2010}
\begin{eqnarray}
S^{\nu}_{\phantom{\nu}\lambda}&\equiv&
\delta^{\,\nu}_{\;\lambda}\,{\mbox{\tt D}}
+ k^{\nu}\,k_{\lambda}\,\zeta^2 + \zeta^{\,\nu}\,\zeta_{\lambda}\,k^2\no&&
- \zeta\cdot k\,(\zeta_{\lambda}\,k^{\nu} + \zeta^{\nu}\,k_{\,\lambda});
\end{eqnarray}
\[
{\mbox{\tt D}}\;\equiv\;(\zeta\cdot k)^2-\zeta^2\,k^2\;=\;\textstyle\frac12\;S^{\nu}_{\phantom{\nu}\nu}.
\]
With a help of the latter equality one can find that
\begin{equation}
S^{\,\mu\lambda}\,\varepsilon_{\lambda\nu\alpha\beta}\,\zeta^\alpha k^{\beta}\;
=\;{\mbox{\tt D}}\;\varepsilon^{\,\mu}_{\phantom{\mu}\nu\alpha\beta}\,\zeta^\alpha k^{\beta} .
\end{equation}
Then to our purpose it is convenient to introduce two orthonormal, one-dimensional, Hermitian projectors
\begin{eqnarray}
\bfpi^{\,\mu\nu}_{\,\pm}&\equiv&
\frac{S^{\,\mu\nu}}{2\,{\mbox{\tt D}}}\;
\pm\;\frac{i}{2}\,
\varepsilon^{\mu\nu\alpha\beta}\,\zeta_{\alpha}\,k_{\beta}\,{\mbox{\tt D}}^{\,-\frac12}\no&&
=\left(\bfpi^{\,\nu\mu}_{\,\pm}\right)^\ast=\left(\bfpi^{\,\mu\nu}_{\,\mp}\right)^\ast;
\quad(\mbox{\tt D}>0) . \label{projectors1}
\end{eqnarray}
A couple of chiral polarization vectors for the MCS field can be constructed out of constant tetrades  $\epsilon_\nu$
\begin{eqnarray}
 \varepsilon_{\,\pm}^{\,\mu\ast}(k)\equiv \bfpi^{\,\mu\lambda}_{\,\pm}\,\epsilon_\mu . \label{projectors2}
\end{eqnarray}
Their properties were thoroughly described  in \cite{AACGS2010}.

In order to obtain the normal modes of propagation of the MCS field, let us introduce the
kinetic $4\times4$ Hermitian matrix $ K$ with elements
\begin{eqnarray}
&&K_{\,\lambda\nu}\equiv
g_{\,\lambda\nu}(k^2-m^2_t) +
i\varepsilon_{\lambda\nu\alpha\beta}\,\zeta^\alpha k^{\beta};\no&&
K_{\,\lambda\nu}=K^{\,\ast}_{\,\nu\lambda} .\end{eqnarray}
We obtain the general solution for $\zeta\cdot x<0$ from the relations (\ref{projectors1}), (\ref{projectors2})
\begin{eqnarray}
&&K^{\,\mu}_{\phantom{\mu}\nu}\,\varepsilon^{\,\nu}_{L}(k) =(k^2-m^2_t) \,
\varepsilon^{\,\mu}_{L}(k),\nonumber\\
&&K^{\,\mu}_{\phantom{\mu}\nu}\,\varepsilon^{\,\nu}_{\pm}(k)\no&& =
\left[\,\delta^{\,\mu}_{\phantom{\mu}\nu}(k^2-m^2_t)
+ {\sqrt{\mbox{\tt D}}}\,\left(\,\bfpi^{\,\mu}_{\,+\,\nu}\;-\;\bfpi^{\,\mu}_{\,-\,\nu}\,\right)\,\right]
\varepsilon^{\,\nu}_{\pm}(k)\nonumber\\
&&= \left(k^2-m^2_t \pm\,\sqrt{\mbox{\tt D}} \,\right)\,
\varepsilon^{\,\mu}_{\pm}(k) .
\end{eqnarray}

\section{Solution of modified Dirac equation} \label{appDiracEq}
We start from the Dirac equation written in momentum phase space:
\ba
(\gamma^\mu p_\mu - m - \gamma^\mu b_\mu \gamma^5)\psi = 0;
\ea
It is convenient to consider expression:
\ba
&&\gamma^0 \gamma^1 \gamma^5=-i \gamma^0 \gamma^1 \gamma^0 \gamma^1 \gamma^2 \gamma^3\no&& = - i \gamma^2 \gamma^3=-\left(
                    \begin{array}{cc}
                      \sigma_1 & 0 \\
                      0 & \sigma_1 \\
                    \end{array}
                  \right) \equiv -  \hat \sigma_1
;
\ea
Using this expression and multiplying the first equation by $\gamma^0$ we get,
\ba
(\gamma^0 \gamma^\mu p_\mu - \gamma^0 m + b \hat \sigma_1) \psi=0.
\label{eqWithGamma}
\ea
Now we introduce two projectors,
\ba
\mathbf {P_{\pm}}\equiv \frac{I\pm\hat\sigma_1}{2}; \qquad \psi_\pm = \mathbf {P_\pm} \psi.
\ea
One can use these projectors to simplify (\ref{eqWithGamma}) with a help of the next expressions,
\ba
&&[\hat \sigma_1, \gamma_0]=0; \quad [\hat \sigma_1, \gamma_1]=0;\no&& \{\hat \sigma_1, \gamma_2\}=0; \quad \{\hat \sigma_1, \gamma_3\}=0,
\ea
namely,
\ba
&&\mathbf {P_+}(\gamma^0 \gamma^\mu p_\mu - \gamma^0 m + b \hat \sigma_1) \psi=0 \Longleftrightarrow \nonumber \\
&&(p_0 - \alpha_1 p_1 -\gamma^0 m + b) \psi_+ \no&&- \alpha_\bot p_\bot \psi_-=0;
\label{eq-}
\ea
and
\ba
&&(p_0 - \alpha_1 p_1 -\gamma^0 m - b) \psi_-\no&& - \alpha_\bot p_\bot \psi_+=0;
\label{eq+}
\ea
where $\alpha_i = \gamma^0 \gamma^i$, and $\alpha_\bot p_\bot = \alpha_2 p_2 + \alpha_3 p_3$. To solve our equations (\ref{eq-}-\ref{eq+}) we multiply (\ref{eq-}) by $\alpha_\bot p_\bot$ and get,
\ba
\psi_-=\frac{(\alpha_\bot p_\bot)(p_0-\alpha_1 p_1 -\gamma^0 m + b)}{p_\bot^2} \psi_+.
\ea
Using this relation one can easily write,
\ba
&&\left[(p_0 - \alpha_1 p_1 -\gamma^0 m - b)\frac{\alpha_\bot p_\bot}{p_\bot^2}\times\right.\\&&\left.\times(p_0-\alpha_1 p_1 -\gamma^0 m + b) - \alpha_\bot p_\bot\right]\psi_+ = 0;\nonumber
\ea
and for $\psi_-, \psi_+$ get equations,
\ba
&&(p_0^2 - \mathbf{p^2} - m^2 - b^2 \no&&\pm 2b (\alpha_1 p_1 + \gamma^0 m) )\psi_\pm=0
\label{eq+-}
\ea
To solve this equation we express $\psi_\pm$ as $\psi_\pm=\left(
                                                    \begin{array}{c}
                                                      \phi_\pm \\
                                                      \xi_\pm \\
                                                    \end{array}
                                                  \right)$.
Using new notation we rewrite  (\ref{eq+-}) as,
\ba
&&(p_0^2 - \mathbf{p^2} - m^2 \no&&- b^2 \pm 2b m)\phi_\pm \pm 2b p_1 \sigma_1 \xi_\pm =0\\
&&(p_0^2 - \mathbf{p^2} - m^2 \no&&- b^2 \mp 2b m)\xi_\pm \pm 2b p_1 \sigma_1 \phi_\pm =0
\ea
From these equations it is easy to get,
\ba
&&\left[(p_0^2 - \mathbf{p^2} - m^2 - b^2)^2 \right.\no&&\left.- 4 b^2 m^2 - 4 b^2 p_1^2 \right] \phi_\pm=0,
\ea
wherefrom the dispersion law is,
\ba
p_0^2=\mathbf{p^2} + m^2 + b^2 \pm 2b \sqrt{m^2 + p_1^2}.
\ea
Thus the Dirac field $\psi(x)$ may be written in form,
\ba
&&\psi(x) = \sum_{A=\pm} u_A(p) e^{-i\hat p \hat x + p_{1A} x_1}, \ea where,\ba &&p_{1\mp}=\no
&&\!\!\sqrt{p_0^2 - p_\bot^2+3b^2-m^2 \pm 2b\sqrt{p_0^2- p_\bot^2+2b^2-2m^2}}.\nonumber
\ea

\section{Transmission/reflection through the boundary }\label{transmission}
Equation (\ref{systspat}) in a half-space $x_1>0$ enforce $v$ to satisfy the following conditions
\begin{eqnarray}
\left\lbrace
\begin{array}{l}
\tilde v_{{2+}_{\overrightarrow{\leftarrow}}}=\frac{k_2k_3+i\omega\sqrt{\omega^2-k_\bot^2}}{\omega^2-k_2^2}\tilde v_{{3+}_{\overrightarrow{\leftarrow}}}; \\
\tilde v_{{2-}_{\overrightarrow{\leftarrow}}}=\frac{k_2k_3-i\omega\sqrt{\omega^2-k_\bot^2}}{\omega^2-k_2^2}\tilde v_{{3-}_{\overrightarrow{\leftarrow}}};
\\
\tilde v_{{0+}_{\overrightarrow{\leftarrow}}}=-\frac{\omega k_3 - i k_2 \sqrt{\omega^2-k_\bot^2}}{2(\omega^2-k_\bot^2)}\tilde v_{{3+}_{\overrightarrow{\leftarrow}}};\\
\tilde v_{{0-}_{\overrightarrow{\leftarrow}}}=-\frac{\omega k_3 + i k_2 \sqrt{\omega^2-k_\bot^2}}{2(\omega^2-k_\bot^2)}\tilde v_{{3-}_{\overrightarrow{\leftarrow}}};\\
\tilde v_{{2L}_{\overrightarrow{\leftarrow}}}=\frac{k_2}{k_3}\tilde v_{{3L}_{\overrightarrow{\leftarrow}}}; \\
\tilde v_{{0L}_{\overrightarrow{\leftarrow}}}=-\frac{\omega}{k_3}\tilde v_{{3L}_{\overrightarrow{\leftarrow}}}.\\
\end{array}\right.
\label{v}
\end{eqnarray}
Thus we have the solutions in both half-spaces, and we should now match them on the boundary. If we believe that all contribution to the vector field $A$ are continuous, the integration over $x_1$ from $-\varepsilon$ to $\varepsilon$ will give us the next relations \cite{AK}
 \begin{eqnarray}
\left\lbrace
\begin{array}{ll}
\tilde u_{0 \rightarrow}^{(L)}=&\frac{\omega^2}{\omega^2-k_{\bot}^2}\tilde u_{0 \rightarrow}+\frac{\omega k_3}{\omega^2-k_{\bot}^2}\tilde u_{3 \rightarrow} + \frac{\omega k_2}{\omega^2-k_{\bot}^2}\tilde u_{2 \rightarrow};\nonumber\\
\tilde u_{0 \rightarrow}^{(\pm)}=&-\frac{k_{\bot}^2}{2(\omega^2-k_{\bot}^2)}\tilde u_{0 \rightarrow}-\frac{\omega k_3\mp i k_2 \sqrt{\omega^2 - k_{\bot}^2}}{2(\omega^2-k_{\bot}^2)}\tilde u_{3 \rightarrow}\\&-\frac{\omega k_2\pm i k_3 \sqrt{\omega^2 - k_{\bot}^2}}{2(\omega^2-k_{\bot}^2)}\tilde u_{2 \rightarrow}.
\label{u0}
\end{array}\right.
\end{eqnarray}
\begin{eqnarray}
\left\lbrace
\begin{array}{ll}
\tilde u_{2 \rightarrow}^{(L)}=&-\frac{k_2^2}{\omega^2-k_{\bot}^2}\tilde u_{2 \rightarrow}-\frac{\omega k_2}{\omega^2-k_{\bot}^2}\tilde u_{0 \rightarrow}\\& - \frac{k_2 k_3}{\omega^2-k_{\bot}^2}\tilde u_{3 \rightarrow};\nonumber\\
\tilde u_{2 \rightarrow}^{(\pm)}=&\frac{\omega^2 - k_3^2}{2(\omega^2-k_{\bot}^2)}\tilde u_{2 \rightarrow}+\frac{\omega k_2\mp i k_3 \sqrt{\omega^2 - k_{\bot}^2}}{2(\omega^2-k_{\bot}^2)}\tilde u_{0 \rightarrow}\\&+\frac{k_2 k_3\mp i \omega \sqrt{\omega^2 - k_{\bot}^2}}{2(\omega^2-k_{\bot}^2)}\tilde u_{3 \rightarrow}.
\end{array}\right.
\end{eqnarray}
\begin{eqnarray}
\left\lbrace
\begin{array}{ll}
\tilde u_{3 \rightarrow}^{(L)}=&-\frac{k_3^2}{\omega^2-k_{\bot}^2}\tilde u_{3 \rightarrow}-\frac{\omega k_3}{\omega^2-k_{\bot}^2}\tilde u_{0 \rightarrow}\\ &- \frac{k_2 k_3}{\omega^2-k_{\bot}^2}\tilde u_{2 \rightarrow};\nonumber\\
\tilde u_{3 \rightarrow}^{(\pm)}=&\frac{\omega^2 - k_2^2}{2(\omega^2-k_{\bot}^2)}\tilde u_{3 \rightarrow}+\frac{\omega k_3\pm i k_2 \sqrt{\omega^2 - k_{\bot}^2}}{2(\omega^2-k_{\bot}^2)}\tilde u_{0 \rightarrow}\\ &+\frac{k_2 k_3\pm i \omega \sqrt{\omega^2 - k_{\bot}^2}}{2(\omega^2-k_{\bot}^2)}\tilde u_{2 \rightarrow} .
\label{u3}
\end{array}\right.
\end{eqnarray}
where $\tilde u_{\nu \rightarrow}=\sum_{A=L,\pm} \tilde u_{\nu \rightarrow}^{(A)}$. Each component of
the incoming amplitudes has its own transmission coefficient \cite{AK},
\ba
\tilde v_{\nu A \rightarrow}=\frac{2k_{1}^0}{k_{1}^0+k_{1A}^{CS}}\tilde u_{\nu \rightarrow}^{(A)}.
\label{tr}
\ea

\section{Boundary effects for electrons} \label{boundaryElectrons}
If we are interested in electrons escaping the parity-breaking domain, we use the next solutions,
\ba
\psi_1(x)&=& \sum_{A=\pm} u_{A\rightarrow}(p) e^{-i\hat p \hat x + p_{1A} x_1} \no &&+ \sum_{A=\pm} u_{A\leftarrow}(p) e^{-i\hat p \hat x - p_{1A} x_1};
\label{solution1View} \\
\psi_2(x)&=& w(p) e^{-i\hat p \hat x + p_{10} x_1}.
\label{solution2View}
\ea
Here the first term in $\psi_1$ stands for the falling electrons, the second term is for the reflected ones. $\psi_2$ describes electrons penetrating through the boundary and propagate in accordance to the Dirac equation with  $p_{10}=\sqrt{p_0^2-p_\bot^2-m^2}$.
The Dirac equation of our system may be used to obtain matching conditions, we consider the small area near the boundary,
\ba
(i \gamma^\mu \partial_\mu - m -\gamma^\mu b_\mu \gamma_5)\psi |_{x_1=-\varepsilon}^{x_1=+\varepsilon}=0
\ea
Using our choice of $b$-vector, we get,

\ba
(i \gamma^1 \partial_1 - \gamma^1 b \theta(-x_1)\gamma^5)\psi|_{-\varepsilon}^{+\varepsilon} = 0;
\ea
Continuity of $\psi$ is the first matching condition. Since $\psi$ is continuous, previous expression leads to,
\ba
\partial_1\psi|_{x_1=+\varepsilon}-\partial_1\psi|_{x_1=-\varepsilon}=-i b \gamma^5 \psi|_{x_1=0}.
\ea
We now use the (\ref{solution1View}, \ref{solution2View}) forms of solutions and rewrite matching conditions in components,
\ba
p_{1A}(u_{A0\leftarrow}-u_{A0\rightarrow})-p_{10}w_{A0}=b w_{A2}; \nonumber\\
p_{1A}(u_{A1\leftarrow}-u_{A1\rightarrow})-p_{10}w_{A1}=b w_{A3}; \nonumber \\
p_{1A}(u_{A2\leftarrow}-u_{A2\rightarrow})-p_{10}w_{A2}=b w_{A0}; \nonumber \\
p_{1A}(u_{3A\leftarrow}-u_{A3\rightarrow})-p_{10}w_{A3}=b w_{A1}. \nonumber
\ea
or, since $w_\mu = u_{\rightarrow\mu} + u_{\mu \leftarrow}$,
\ba
u_{A0\leftarrow}=\frac{2b p_{1A} u_{A2\rightarrow} - (p_{1A}^2 - p_{10}^2 + b^2)u_{A0\rightarrow}}{b^2+(p_{1A}+p_{10})^2};\no
u_{A1\leftarrow}=\frac{2b p_{1A} u_{A3\rightarrow} - (p_{1A}^2 - p_{10}^2 + b^2)u_{A1\rightarrow}}{b^2+(p_{1A}+p_{10})^2};\no
u_{A2\leftarrow}=\frac{2b p_{1A} u_{A0\rightarrow} - (p_{1A}^2 - p_{10}^2 + b^2)u_{A2\rightarrow}}{b^2+(p_{1A}+p_{10})^2};\no
u_{A3\leftarrow}=\frac{2b p_{1A} u_{A1\rightarrow} - (p_{1A}^2 - p_{10}^2 + b^2)u_{A3\rightarrow}}{b^2+(p_{1A}+p_{10})^2}.\nonumber
\ea

To give some quantitative effects let us try to describe the process of reflection of electrons. We deal with wave functions $\psi$, however, only $|\psi|^2$ has the physical meaning. Let us for simplicity take the propagating electron with \[u_{\mu \rightarrow} = (0, u_1, 0, 0)\] and assume that this particle is on the mass-shell $p_1=p_{1A}(\hat p)$. In this case after the reflection of the boundary we would get {\small \[u_{\mu \leftarrow} = (0, \frac{ - (p_{1A}^2 - p_{10}^2 + b^2)u_{1}}{b^2+(p_{1A}+p_{10})^2},0, \frac{2b p_{1A} u_{1}}{b^2+(p_{1A}+p_{10})^2})\]} and
\ba
\!\!\frac{|\psi|^2_{reflected}}{|\psi|^2_{initial}} = \frac{4 b^2 p_{1A}^2 + (p_{1A}^2-p_{10}^2+b^2)^2}{(b^2+(p_{1A}+p_{10})^2)^2}
\ea

\end{document}